\magnification=\magstep1
\hfuzz=6pt 
\baselineskip=16pt
$ $

\vskip 1in

\centerline{\bf Robustness of Adiabatic Quantum Computing}

\bigskip

\centerline{Seth Lloyd, MIT}

\vskip 1cm
\noindent{\it Abstract:} Adiabatic quantum computation for performing
quantum computations such as Shor's algorithm is protected
against thermal errors by an energy gap of size $O(1/n)$, where $n$ is 
the length of the computation to be performed.  

\vskip 1in

Adiabatic quantum computing is a novel form of quantum information
processing that allows one to find solutions to NP-hard problems
with at least a square root speed-up [1].  In some cases, adiabatic
quantum computing may afford an exponential speed-up over classical
computation.  It is known that adiabatic quantum computing is no
stronger than conventional quantum computing, since a quantum computer
can be used to simulate an adiabatic quantum computer.  
Aharonov {\it et al.} showed that adiabatic quantum computing is no
weaker than conventional quantum computation [2].  This paper presents
novel models for adiabatic quantum computation and shows that
adiabatic quantum computation is intrinsically protected against
thermal noise from the environment.  Indeed, thermal noise can
actually be used to `help' an adiabatic quantum computation along. 

A simple way to do adiabatic versions of `conventional' quantum computing
is to use the Feynman pointer consisting of a line of qubits [3].
The Feynman Hamiltonian is
$$ H = - \sum_{\ell=0}^{n-1} U_\ell \otimes |\ell+1\rangle
\langle \ell| + U_\ell^\dagger \otimes |\ell\rangle \langle \ell +1|, 
\eqno(1)$$  
where $U_\ell$ is the unitary operator for the $\ell$'th gate
and $|\ell\rangle$ is a state of the pointer where the $\ell$'th
qubit is 1 and all the remaining qubits are 0.
Clearly, $H$ is local and each of its terms
acts on four qubits at once for two-qubit gates.  If we consider
the pointer to be a `unary' variable, then the each of the
terms of $H$ acts on two qubits and the unary pointer variable.

Assume that the computation has been set up so that all qubits
start in the state 0.  The computation takes place and the answer
is placed in an answer register.  Now a long set of steps, say
$n/2$ takes place in which nothing happens.   Then the computation
is undone, returning the system to its initial state at the
$n-1$'th step.  The computational circuit then wraps around and begins again.
The eigenstates of $H$ then have the form 
$$|b,k\rangle = (1/\sqrt n)\sum_{\ell=0}^{n-1} e^{i2\pi k\ell/n}
U_\ell \ldots U_0 |b\rangle\otimes|\ell\rangle, \eqno(2)$$
with eigenvalue $-2\cos 2\pi k/n$, since
$H|b,k\rangle = -(e^{-i 2\pi k/n} + e^{i 2\pi k/n}) |b,k\rangle$.
These can be thought of as momentum eigenstates for the propagation
of the pointer qubit down the chain.  The $|0,k\rangle$ momentum eigenstates
have the nice feature that if you measure the answer register,
the probability of obtaining the answer is $1/2$.

Feynman used this Hamiltonian to set up a traveling state of the
pointer (a Gaussian superposition of the momentum eigenstates) so that the
computation could take place sequentially.  This propagating
state is mathematically equivalent to a coherent quantum walk
down the chain [4-5].  Landauer pointed out that for any realistic
implementation of such a system, imperfections in the Hamiltonian
would result in Anderson localization and the computation would
get `stuck' and fail to propagate to the end [6].  Localization
is a significant problem for quantum walks in general and for
the Feynman quantum computing model in particular.

One can also use Feynman's Hamiltonian to implement the computation 
adiabatically.  In this case, as will be seen, localization does
not affect the computation. 
Let $$H_0 =\sum_j (1- |0\rangle_j\langle 0|) \otimes |\ell=0\rangle
\langle \ell=0| \eqno(3)$$ be the Hamiltonian one of whose ground states is
the state with all the computer qubits equal to zero and the pointer at
the zero spot of the line.   $H_0$ is degenerate: states where
the pointer is at other places also have zero energy, regardless
of the values of the computational qubits.  First,
prepare the system in the state $|00 \ldots 0\rangle |\ell =0\rangle$
by applying the Hamiltonian $\eta H_0 + H_1$, where
$H_1 = -|\ell=0\rangle\langle \ell=0|$
and $\eta$ is a positive term that sets the overall energy scale.

Now gradually turn on an $H$ term while turning off the $H_1$ term: 
the total Hamiltonian
is $\eta H_0 + (1-\lambda) H_1 + \lambda H$.  As $\lambda$ is turned on,
the system goes to its new ground state $|b=0,k=0\rangle$.
It can be verified numerically and analytically [7]
that the minimum energy gap in 
this system occurs at $\lambda =1$: consequently, the minimum
gap goes as $1/n^2$.  In fact,
the energy gap due to the interaction between the $H_1$ and the 
$H$ terms is just the energy gap of the simpler system consisting
just of the chain qubits on their own, confined to the subspace
in which exactly one qubit is 1: that is, it is the energy
gap of a qubit chain with Hamiltonian $(1-\lambda) H_1
+ \lambda H'$, where $H' = -\sum_\ell |\ell + 1\rangle \langle \ell|
+ |\ell \rangle \langle  \ell + 1|$.  
This gap goes as $1/n^2$. 
Accordingly, the amount of time required to perform the adiabatic
passage is polynomial in $n$.

When the adiabatic passage is complete,
the energy gap of the $H$ term on its own goes as
$1/n^2$ from the cosine dependence of the eigenvalues of $H$:
it is also just the energy gap of the simplified system in the
previous paragraph.  This implies that the adiabatic passage can
accurately be performed in a time polynomial in $n$.
Measuring the answer register now gives the answer to the computation
with probability $1/2$.  This is an alternative (and considerably
simpler) demonstration to that of [2] that `conventional' quantum computation 
can be performed efficiently in an adiabatic setting.  

An interesting feature of this procedure is that the adiabatic
passage can be faulty and still work just fine:
all of the energy eigenstates in the $|b=0, k\rangle$ sector 
give the correct answer to the computation with probability $1/2$,
for any $k$.  The real issue is making sure we do not transition to the
$|b\neq 0,k\rangle$ sector.  But the Hamiltonians $H_1$ and
$H$ do not couple to this sector: so in fact, we can perform
the passage non-adiabatically and still get the answer to
the computation.  For example, if we turn off the $H_1$
Hamiltonian very rapidly and turn on the $H$ Hamiltonian
at the same time, the system is now in an equal superposition of
all the $|b=0,k\rangle$ eigenstates.  If we wait for a time
$ \propto n^2$ (corresponding to the inverse of the minimum
separation between the eigenvalues of $H$), then the state of
the system will be spread throughout the $|b=0,k\rangle$ sector,
and we can read out the answer with probability $1/2$.
This method effectively relies on dispersion of the wavepacket
to find the answer.

Since coherence of the pointer doesn't matter, we can also
apply a Hamiltonian to the pointer that tilts the energy
landscape so that higher pointer values have lower energy.  
(E.g., $H_{\rm pointer} = -\sum_\ell \ell E |\ell\rangle
\langle \ell|$.)  Starting the pointer off in the initial
state above and letting it interact with a thermal environment
will obtain the answer to the computation in time of $O(n)$.
Similarly, in the absence of an environment, 
starting the pointer off in a wavepacket with
zero momentum at time 0 and letting it accelerate will
get the answer to the computation in even shorter time.

Clearly, this method is quite a robust way of performing quantum
computation.  Let us look more closely at the sources of this
robustness.  If $\eta$ is big, then there is a separation of energy
of $O(\eta/n)$ between the $|b=0,k\rangle$ sector --- states
which give the correct answer to the computation ---
and the $|b\neq 0,k\rangle$ sector --- states which give the
incorrect answer to the computation.  This is because
$\langle b\neq 0,k|\eta H_0|b \neq 0,k\rangle = \eta/n$.  
This energy gap goes down only linearly in the length of the computation and
can be made much larger than the gap between the ground and first
excited state by increasing $\eta >> 1$.

This second energy gap is very useful: it means 
that thermal excitations with an energy below the gap
will not interfere with obtaining the proper answer.  That is,
this method is intrinsically error-tolerant in the face
of thermal noise.  Indeed, it
is this $O(\eta/n)$ gap that determines how rapidly the
computation can take place rather than the $O(1/n^2)$ gap
between the ground and excited states.  

Of course, the actual errors in a system that realizes the above
scheme are likely to arise from variability in the applied Hamiltonians.
The energy gap arguments for robustness 
only apply to the {\it translational} dynamics of the system
(this is what makes the analysis of the system tractable in
the first place).  That is, errors that affect each $U_\ell$ on
its own are not protected against: but these are the errors that
cause the computation to come out wrong.  Of course, one can
always program the circuit to perform `conventional' robust
quantum computation to protect against such errors.  One must
be careful, however, that errors that entangle the pointer 
with the logical qubits do not contaminate other qubits: 
conventional robust quantum computation protocols will
have to be modified to address this issue.
Farhi {\it et al.} have recently exhibited error correcting
codes for `conventional' adiabatic quantum computation [1] 
that can protect against such computational errors [8].  

The use of error correcting codes to correct the variation
in the $U_\ell$ may well be overkill.
In any system manufactured to implement adiabatic
quantum computing, these errors in the $U_\ell$ are 
essentially deterministic: the $U_\ell$ could in principle
be measured and their variation from their nominal values
compensated for by tuning the local Hamiltonians.  
Because it involves no added redundancy, such an
approach is potentially more efficient than the use of quantum
error correcting codes.   Exactly how to detect and correct
variations in the $U_\ell$ will depend on the techniques
(e.g., quantum dots or superconducting systems)
used to construct adiabatic quantum circuits.

It is also interesting to note that performing quantum computation
adiabatically is intrinsically
more energy efficient than performing a sequence of quantum
logic gates via the application of a series of external pulses.
The external pulses must be accurately clocked and shaped,
which requires large amounts of energy.  In the schemes investigated
here, the internal dynamics of the computer insure that 
quantum logic operations are performed in the proper order,
so no clocking or external pulses need be applied.
The adiabatic technique also avoids the Anderson localization
problem raised by Landauer.

The above construction requires an external pointer and 
four qubit interactions.  One can also set up a pointerless model that 
requires only pairwise interactions between spin 1/2 particles
(compare the following method with the method proposed in reference [9]).
Let each qubit in the computational circuit correspond
to a particle with two internal states.  Let each wire in the circuit
correspond to a mode that can be occupied by a particle.
The $\ell$'th quantum logic gate then corresponds to an
operator $\tilde H_\ell =  A_\ell +  A^\dagger_\ell$,
where $A_\ell$ is an operator that takes two particles from
the two input modes and moves them to the output modes while
performing a quantum logic operation on their two qubits. 
That is,
$$A_\ell = a^\dagger_{out 1}a_{in 1}  a^\dagger_{out 2}a_{in 2}
\otimes U_\ell. \eqno(5)$$  Note that $A_\ell$ acts only when both input
modes are occupied by a qubit-carrying particle.  If we
use the Hamiltonian $\tilde H = \sum_\ell \tilde H_\ell$ 
in place of $H$ in the construction
above, the ground state of this Hamiltonian is a superposition of
states in which the computation is at various stages of completion.
Just as above, measurement on the ground state will reveal the 
answer to the computation with probability $1/2$.

Note that even though the Hamiltonian in equation (5) involves
a product of operators on four degrees of freedom (the internal
degrees of freedom of the particles together with their
positions), it is nonetheless a physically reasonable local Hamiltonian
involving pairwise interactions between spin-1/2 particles.
To simulate its operation using an array of qubits as in
[2] would require four qubit interactions, as
in the pointer model discussed above.
This point is raised here because of the emphasis in the
quantum computing literature on reducing computation to
pairwise interactions between qubits.  Pairwise interactions 
between particles or fields -- i.e., the sort of interactions
found in nature -- may correspond to interactions between more
than two qubits.

Without further restrictions on the form of the quantum logic
circuit, evaluating the energy gap in this particle model is difficult, 
even for the final Hamiltonian $1-\tilde H$.  But we can always
set up the computational circuit in a way that allows the adiabatic
passage to be mapped directly on the Feynman pointer model above.
The method is straightforward: order the quantum logic gates
as above.  Now insert additional quantum logic gates between
each consecutive pair of gates in the original circuit.  
The additional gate inserted between
the $\ell$'th and $\ell+1$'th quantum logic gates couples one
output qubit of the $\ell$'th quantum logic gate with one
input qubit of the $\ell+1$'th gate, and performs a {\it trivial}
operation $U=1$ on the internal qubits of these gates.
The purpose of these gates is ensure that the quantum logic 
operations are performed in the proper sequence.  Effectively,
one of the qubits from the $\ell$'th gate must `tag' one
of the qubits from the $\ell+1$'th gate before the $\ell+1$'th
gate can be implemented.  Accordingly, we call this trick, a
`tag-team' quantum circuit.  

Tag-team quantum circuits are unitarily equivalent to the Feynman
pointer model with an extra, identity gate inserted between each
of the original quantum logic gates.  Accordingly, the spectral
gap for tag-team quantum circuits goes as $1/n^2$ and the quantum
computation can be performed in time $O(poly(n))$.  Just as for
the pointer version of adiabatic quantum computing, the important 
spectral gap for tag-team adiabatic quantum computation is not
the minimum gap, but rather the gap of size $O(\eta/n)$
between the ground-state manifold of `correct' states and
the next higher manifold of `incorrect' states.  Once
again, the existence of this gap is a powerful protection
against errors in adiabatic quantum computation.

The methods described above represent an alternative derivation
of the fact that adiabatic quantum information 
processing can efficiently perform conventional quantum computation.  
The relative simplicity of the derivation from the original Feymnan
Hamiltonian [3] allows an analysis of the robustness of the scheme
against thermal excitations.
Adiabatic implementations of quantum computation are robust against
thermal noise at temperatures below the appropriate energy gap.
The appropriate energy gap is not the minimum gap, which scales
as $n^2$, but the gap between the lowest sector of eigenstates,
which give the correct answer, and the next sector.  This gap
scales as $\eta/n$, where $\eta$ is an energy parameter that
is within the control of the experimentalist.

\vfill
\noindent{\it Acknowledgements:} This work was supported by ARDA,
ARO, DARPA, CMI, and the W.M. Keck foundation.  The author would
like to thank D. Gottesman and D. Nagaj for helpful conversations.

\vfil\eject

\noindent{\it References:} 

\bigskip

\noindent[1] E. Farhi, J. Goldstone, S. Gutmann, {\it Science} {\bf 292}, 472 (2
001).

\noindent[2] 
D. Aharonov, W. van Dam, J. Kempe, Z. Landau, S. Lloyd, O. Regev,
`Adiabatic Quantum Computation is Equivalent to Standard Quantum Computation,'
Proceedings of the 45th Annual IEEE Symposium on Foundations of Computer 
Science (FOCS'04), 42-51  (2004); 
quant-ph/0405098. 

\noindent[3]
R. Feynman, {\it Found Phys.} {\bf 16}, 507-531 (1986).

\noindent[4]
E. Farhi, S. Gutmann, {\it Phys. Rev. A} {\bf 58}, 915-928 (1998).

\noindent[5]
D. Aharonov, A. Ambainis, J. Kempe, U. Vazirani,
{\it Proceeding of the thirty-third annual ACM symposium on
Theory of Computing},  pp. 50-59 (2001).

\noindent[6]
R. Landauer, {\it Phil. Trans.: Phys. Sci. and Eng.}, {\bf 353}, 
367-376 (1995).

\noindent[7] P. Deift, M.B. Ruskai, W. Spitzer,
arXiv:quant-ph/0605156.

\noindent[8]
S.P. Jordan, E. Farhi, P. Shor, {\it Phys Rev A} {\bf 74}, 052322 
(2006); quant-ph/0512170.

\noindent[9] 
    A. Mizel, D.A. Lidar, M. Mitchell,
{\it Phys. Rev. Lett.} {\bf 99}, 070502 (2007) 
arXiv:quant-ph/0609067. 

\vfill\eject\end